\documentclass[aps,floats,twocolumn,superscriptaddress]{revtex4-2}
\usepackage{graphicx}
\usepackage{dcolumn}
\usepackage{enumerate}
\usepackage{color}
\DeclareGraphicsExtensions{.eps,.ps,.pdf}
\usepackage{amsmath,amssymb,bbold,bm}
\usepackage[utf8]{inputenc}
\usepackage{subcaption}
\usepackage[singlelinecheck=off,justification=raggedright]{caption}

\begin{document}

\newcommand{\dea}{d\Tc$/$d$\epsilon_{11}$}
\newcommand{\deb}{d\Tc$/$d$\epsilon_{22}$}
\newcommand{\dei}{d\Tc$/$d$\epsilon_{\rm ii}$}
\newcommand{\Tc}{$T_{\rm c}$}

\title{Effect of pseudogap on electronic anisotropy
in the strain dependence of the superconducting $T_c$ 
of underdoped YBa$_2$Cu$_3$O$_y$}

\author{M. Frachet}
\thanks{Present adress: Institute for Quantum Materials and Technologies, Karlsruhe Institute of Technology, D-76344,
Eggenstein-Leopoldshafen, Germany}
\affiliation{LNCMI-EMFL, CNRS UPR3228, Univ. Grenoble Alpes, Univ. Toulouse, Univ. Toulouse 3, INSA-T,  Grenoble and Toulouse, France}
\author{Daniel J. Campbell}
\affiliation{LNCMI-EMFL, CNRS UPR3228, Univ. Grenoble Alpes, Univ. Toulouse, Univ. Toulouse 3, INSA-T,  Grenoble and Toulouse, France}
\author{Anne Missiaen}
\affiliation{LNCMI-EMFL, CNRS UPR3228, Univ. Grenoble Alpes, Univ. Toulouse, Univ. Toulouse 3, INSA-T,  Grenoble and Toulouse, France}
\author{S. Benhabib}
\affiliation{LNCMI-EMFL, CNRS UPR3228, Univ. Grenoble Alpes, Univ. Toulouse, Univ. Toulouse 3, INSA-T,  Grenoble and Toulouse, France}
\author{Francis Lalibert\'e}
\affiliation{LNCMI-EMFL, CNRS UPR3228, Univ. Grenoble Alpes, Univ. Toulouse, Univ. Toulouse 3, INSA-T,  Grenoble and Toulouse, France}
\author{B. Borgnic}
\affiliation{LNCMI-EMFL, CNRS UPR3228, Univ. Grenoble Alpes, Univ. Toulouse, Univ. Toulouse 3, INSA-T,  Grenoble and Toulouse, France}
\author{T. Loew}
\affiliation{Max-Planck-Institut für Festkörperforschung, Heisenbergstrasse 1, Stuttgart, D-70569, Germany}
\author{J. Porras}
\affiliation{Max-Planck-Institut für Festkörperforschung, Heisenbergstrasse 1, Stuttgart, D-70569, Germany}
\author{S. Nakata}
\affiliation{Max-Planck-Institut für Festkörperforschung, Heisenbergstrasse 1, Stuttgart, D-70569, Germany}
\author{B. Keimer}
\affiliation{Max-Planck-Institut für Festkörperforschung, Heisenbergstrasse 1, Stuttgart, D-70569, Germany}
\author{M. Le Tacon}
\affiliation{Institute for Quantum Materials and Technologies, Karlsruhe Institute of Technology, D-76344,
Eggenstein-Leopoldshafen, Germany}
\author{Cyril Proust}
\affiliation{LNCMI-EMFL, CNRS UPR3228, Univ. Grenoble Alpes, Univ. Toulouse, Univ. Toulouse 3, INSA-T,  Grenoble and Toulouse, France}
\author{I. Paul}
\thanks{indranil.paul@univ-paris-diderot.fr}
\affiliation{Laboratoire Matériaux et Phénomènes Quantiques, CNRS, Université de Paris, F-75205 Paris, France}
\author{David LeBoeuf}
\thanks{david.leboeuf@lncmi.cnrs.fr}
\affiliation{LNCMI-EMFL, CNRS UPR3228, Univ. Grenoble Alpes, Univ. Toulouse, Univ. Toulouse 3, INSA-T,  Grenoble and Toulouse, France}

\date{\today}

\begin{abstract}
For orthorhombic superconductors we define thermodynamic anisotropy 
$N \equiv d T_c/d \epsilon_{22} - dT_c/d \epsilon_{11}$ as the difference in how superconducting $T_c$
varies with strains $\epsilon_{ii}$, $i=(1, 2)$,
along the in-plane directions. We study the hole doping ($p$) dependence of $N$ 
on detwinned single crystals of underdoped YBa$_2$Cu$_3$O$_y$ (YBCO) using ultrasound technique.
While the structural orthorhombicity of YBCO reduces monotonically with decreasing 
doping over $0.065 <p<0.16$, we find that the thermodynamic anisotropy shows an intriguing enhancement at intermediate doping level, which is of electronic origin. 
Our theoretical analysis shows that the enhancement of the electronic anisotropy 
can be related to the pseudogap potential in the electronic specturm
that itself increases when the Mott insulating state is approached. 
Our results imply that the pseudogap is controlled by a local energy scale that can be tuned by 
varying the nearest neighbor Cu-Cu bond length. Our work opens the possibility to strain engineer 
the pseudogap potential to enhance the superconducting \Tc.

\end{abstract}

\maketitle


The link between electronic anisotropy and high temperature superconductivity in the cuprates and the iron based 
systems is a subject of great current interest. While a lot of progress on this topic has been made for 
the iron based systems, relatively less is known about the in-plane electronic anisotropy observed in the 
pseudogap state of certain underdoped 
cuprates \cite{ando02,sato17,hinkov08,Mangin-Thro17,Achkar16,daou10,cyrchoi15,lawler10,zheng17,Wu15,ishida20,Wu17}.
The microscopic factors governing this anisotropy are currently 
unknown, and are the subject of intense research \cite{nie14,tranquada15,morice18,sachdev19,orth19,gull09}. 
Evidently, identifying the source of this anisotropy is of utmost importance
for understanding the pseudogap state and the phase diagram of the cuprates. The
purpose of the current joint experimental and theoretical study is to address this issue.

Experimentally, the anisotropy has been probed using a variety of techniques including 
in-plane electrical conductivity \cite{ando02}, torque magnetometry \cite{sato17},
neutron \cite{hinkov08,Mangin-Thro17} and X-ray \cite{Achkar16} diffraction,  
Nernst coefficient \cite{daou10,cyrchoi15}, scanning tunneling spectroscopy \cite{lawler10,zheng17}, 
nuclear magnetic resonance \cite{Wu15}, and elastoresistivity \cite{ishida20}. 
One school of thought has identified the pseudogap temperature $T^{\star}$ with an electronic nematic phase transition \cite{sato17}. However, the situation is unclear because signatures of diverging nematic correlation, expected near a nematic phase transition~\cite{gallais13}, have
not been detected in electronic Raman response in Bi$_2$Sr$_2$CaCu$_2$O$_{8+\delta}$~\cite{auvray19}.

\begin{figure}
    \includegraphics[width=9cm]{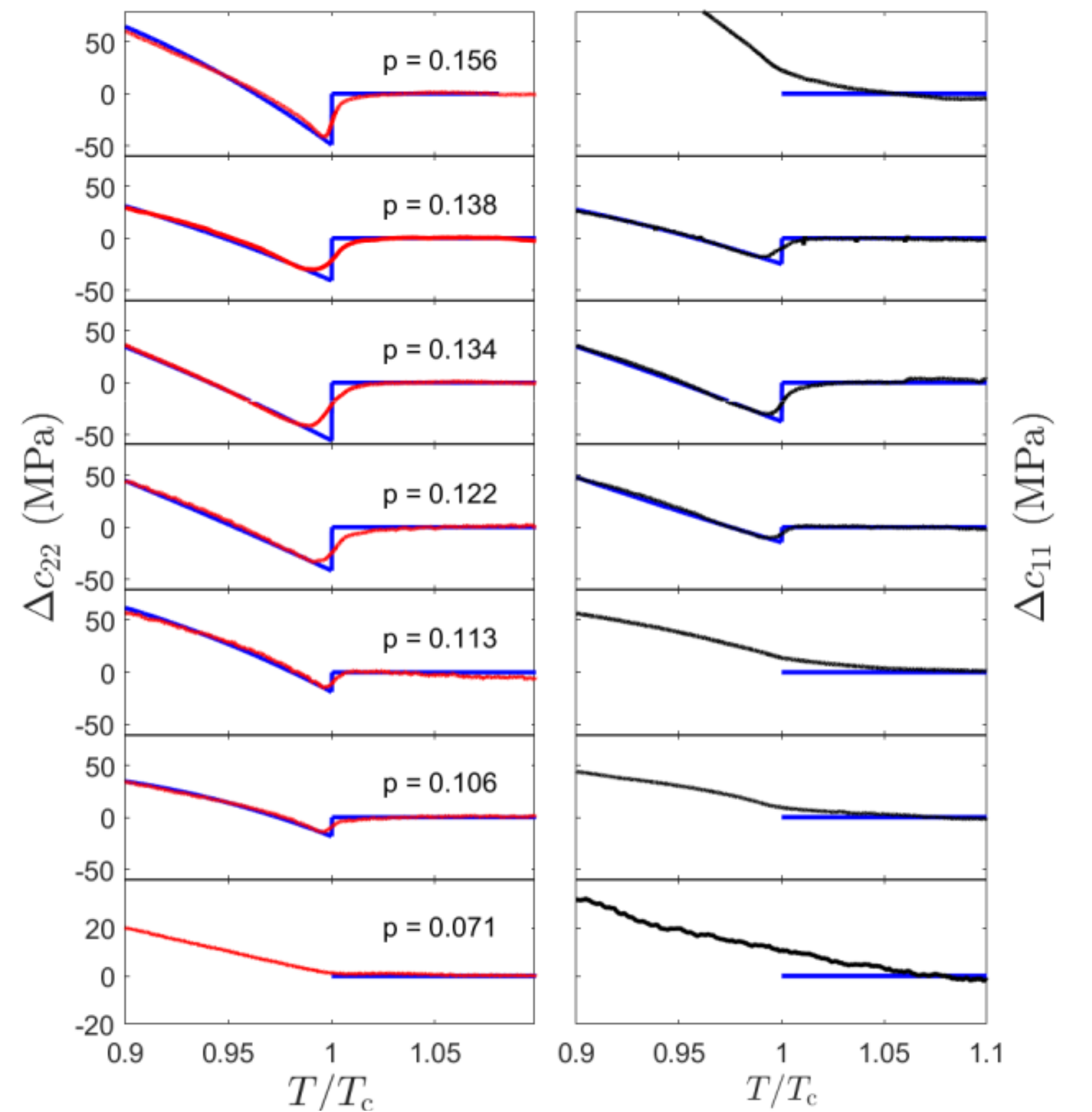}
    \caption{Superconducting contribution to $c_{22}(T)$ (red, left column) and $c_{11}(T)$ (black, right column) near \Tc~as a function of doping in YBCO. A fit based on a thermodynamic model \cite{SM} is shown in blue. It is used to extract $\Delta c_{\rm ii}(T_{\rm c})$, the mean-field jump-like anomaly at \Tc. When no jump is observed we can extract an upper limit for \dei~which depends on measurement noise level and on the amplitude of the specific heat jump at \Tc. \Tc~is defined as the position of the mean-field anomaly in  $\Delta c_{\rm ii}(T)$. The scale is the same for all doping levels except for $p=0.071$ where the vertical scale is reduced for clarity.}
    \label{fig:SC}
\end{figure}

Motivated by the status quo, we study the doping evolution
of the thermodynamic anisotropy
$N \equiv dT_{\rm c}/d \epsilon_{22} - dT_{\rm c}/d \epsilon_{11}$, where $dT_{\rm c}/d \epsilon_{ii}$ is the variation
of the superconducting \Tc~with uniaxial strain $\epsilon_{ii}$,  $ii = (11, 22)$, of underdoped
YBa$_2$Cu$_3$O$_y$ (YBCO).
The experimental technique involves measuring the jumps in the associated elastic
constants $\Delta c_{ii}$ at \Tc~using sound velocity measurements (see Fig.~\ref{fig:SC}),
from which we extract
$dT_{\rm c}/d \epsilon_{ii}$ using the Ehrenfest relationship.  The advantage of this method is that the
strain dependence of \Tc~is obtained in \emph{zero} applied static strain, as explained below. 
Consequently, the measurement is free of nonlinear effects that can be difficult to interpret.
To the best of our knowledge, such strain dependence of \Tc~has not been reported earlier in YBCO.
This thermodynamic anisotropy is in line with earlier studies of uniaxial pressure dependencies of \Tc~\cite{kraut93,pasler00}.
However, converting them into strain dependencies is difficult due to
the large uncertainties in the experimental values of the elastic constant tensor.

\begin{figure}
\center
    \includegraphics[width=8cm]{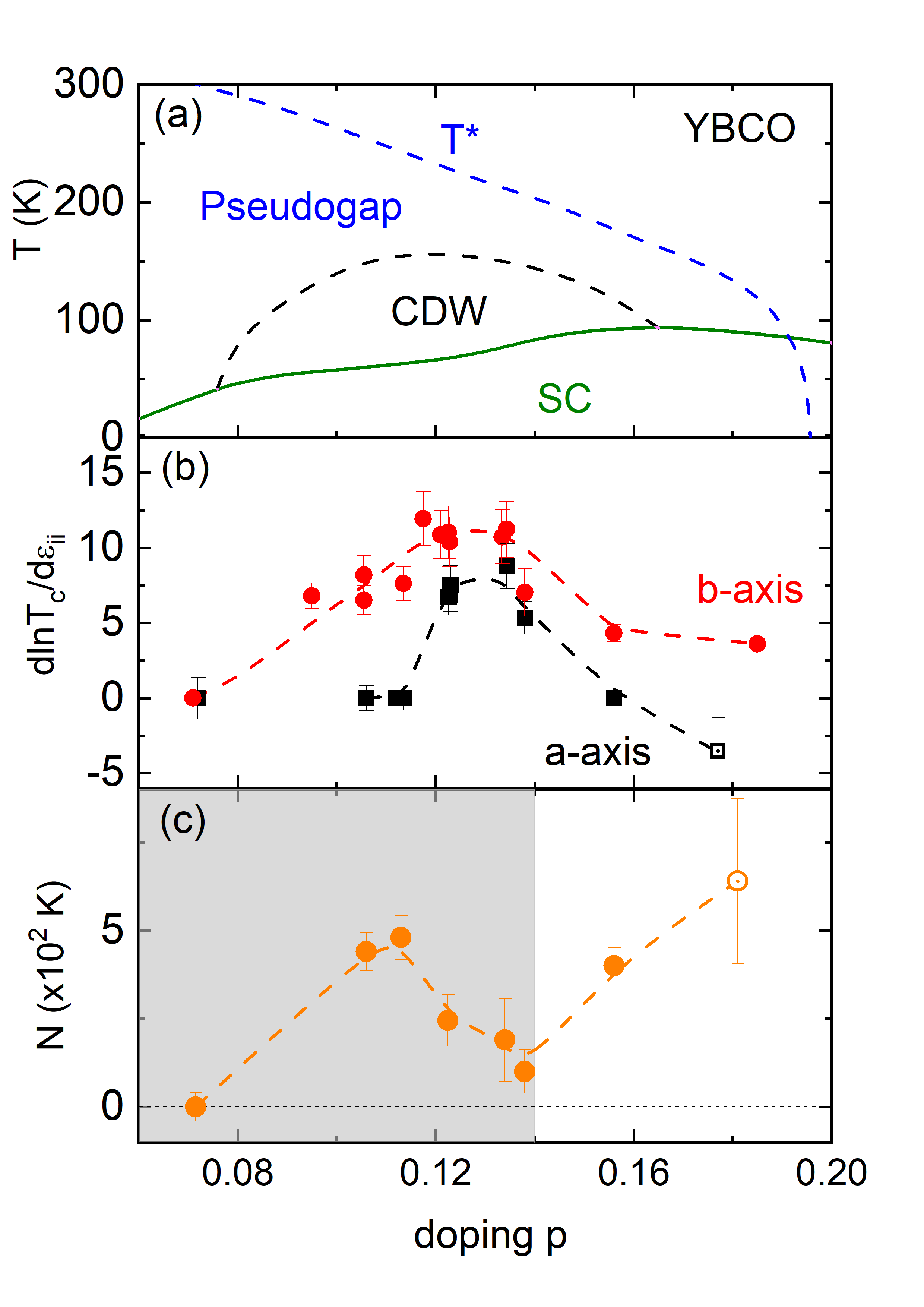}
    \caption{a) Temperature - doping phase diagram of YBCO in zero magnetic field. Green line is the superconducting dome, black dashed line is the dome of short range CDW, blue dashed line is the pseudogap onset temperature $T^\star$. b) Doping dependence of 
$d \ln T_c/d \epsilon_{11}$ (black) and $d \ln T_c/d \epsilon_{22}$ (red). c) Thermodynamic anisotropy $N=~$\deb$-$\dea. The shaded area highlights the doping range where the anisotropy is mostly controled 
by the physics of the CuO$_2$ planes, and consequently where comparison with the theoretical model is most relevant (see text). Dashed lines are guide to the eyes. Data from this study are shown using solid symbols \cite{footnote}. }
\label{fig:diag}
\end{figure}

Our main observation is that, while the crystalline anisotropy, namely the orthorhombicity,
reduces monotonically with decreasing hole doping 
over $0.065 < p < 0.16$ \cite{jorgensen90,casalta96,kruger97}, the thermodynamic anisotropy
$N(p)$ is a non monotonic function of $p$ (see Fig.~\ref{fig:diag}). In particular,
in the range $0.11 < p < 0.14$, $N(p)$ does not track the orthorhombicity, instead
it increases when $p$ is reduced. We therefore conclude that the observed non-monotonic evolution is rooted in electronic effects.
Our theoretical modeling suggests that the enhanced electronic anisotropy in this doping range
is due to the pseudogap and its doping dependence. In
other words, the increase in anisotropy with decreasing doping level reflects
the fact that the pseudogap potential enhances
as the system approaches the Mott insulating state by reducing $p$.

\begin{table}[h]
    \begin{center}
        \begin{tabular}{lcccc}
\hline

$~~y~$~~ & ~~$T_{\rm c}$~(K)~ &    $~~~p$~(holes/Cu) & \dea (K) & \deb (K)\\

\hline
\hline

6.45     & 34.0   & 0.071  & $0 \pm 50$ & $0 \pm 50$\\
6.48 & 55.8 & 0.095 & - & $380 \pm 52$\\
6.51 & 60.0 & 0.106 & $0 \pm 50$ & $440 \pm 70$\\
6.55         & 62.5   & 0.113 & $0 \pm 50$ & $480 \pm 76$  \\
6.67         & 67.7   & 0.122  & $475 \pm 85$ & $720 \pm 115$ \\
6.75         & 77.0   & 0.134   & $655 \pm 135$ & $845 \pm 175$\\
6.79 & 82.0 & 0.138 & $450 \pm 102$ & $ 560 \pm 100 $\\
6.87     & 92.3   & 0.156   & $0 \pm 50$ & $400 \pm 65$\\
6.99    &  88.5  &  0.185  & - & $320 \pm 34$\\

\hline
\hline

    \end{tabular}
    \end{center}
\caption{Characteristics of the YBCO samples measured in this study:
the oxygen content $y$;
the superconducting transition temperature in zero magnetic field \Tc;
the hole concentration (doping) $p$, obtained from \Tc~\cite{liang06}. Typical \dea~and \deb~are given for each oxygen content $y$.}
\label{tab:tab1}
\end{table}

The sound velocities
of several detwinned YBCO samples (see Table \ref{tab:tab1} for characteristics) measured across their superconducting transition temperature \Tc~are shown
in Fig. \ref{fig:SC} (see Supplementary Material (SM) \cite{SM} for experimental details and additional data).
We focus on the elastic constants $c_{11}$ and $c_{22}$ corresponding to longitudinal modes with propagation along the $a$-axis and $b$-axis of the orthorhombic crystal structure of YBCO, respectively.

In Fig. \ref{fig:SC} we show the superconducting contribution to the elastic constants, obtained after subtraction of the thermally activated anharmonic background \cite{varshni70}. The latter consists of a change of slope and curvature below \Tc~and a downward,
mean-field jump $\Delta c_{\rm ii}(T_{\rm c})$ at \Tc. This jump is a consequence of having a term
$\phi^2 \epsilon_{ii}$ in the free energy that couples the strain with the superconducting order
parameter $\phi$~\cite{SM}.
Here we focus on the magnitude of this jump $\Delta c_{\rm ii}(T_{\rm c})$,
which strongly depends on doping level and on propagation direction.
In particular, an anisotropy is observed between $\Delta c_{11}(T_{\rm c})$ and $\Delta c_{22}(T_{\rm c})$ at $p\leq0.11$ and  $p\geq0.156$: at $T=T_{\rm c}$,~a clear jump is observed in $\Delta c_{22}(T)$ but no jump is observed in $\Delta c_{11} (T)$. However, at intermediate doping level the anisotropy is reduced, with a clear jump resolved in both modes. The magnitude of $\Delta c_{\rm ii}(T_{\rm c})$ is governed by 
the Ehrenfest relationship \cite{SM,testardi71,rehwald73,millis88,luthi}
\begin{equation}
\label{eq:exp-1}
\Delta c_{\rm ii}(T_{\rm c})=-\frac{\Delta C_{\rm p}(T_{\rm c})} {T_{\rm c}}\frac{1}{V_m}\Big(\frac{dT_{\rm c}}{d\epsilon_{\rm ii}}\Big)^2,
\end{equation}
with $\Delta C_{\rm p}(T_{\rm c})$ the jump in the heat capacity at \Tc, and $V_m$ the molar volume. Thus, the anisotropy in $\Delta c_{ii}$ implies a difference between \dea~and \deb.
We use a thermodynamic model to fit  the data in Fig. \ref{fig:SC} and to extract $\Delta c_{\rm ii}(T_{\rm c})$ \cite{nohara95, SM}. We then use Eq. \ref{eq:exp-1}, in combination with specific heat \cite{junod89,wuhl91,claus92,loram01,marcenat15} and uniaxial pressure dependence of \Tc~data \cite{meingast90,meingast91,kraut93,welp92,welp94,ludwig96,barber21} in order to determine the amplitude and sign of \dei~respectively.
Since the acoustic waves are nothing but strain waves at finite frequency and wavevector, 
our method allows the extraction of $d T_c/d \epsilon_{ii}$ without actually applying static uniform 
strain $\epsilon_{ii}$~\cite{SM}.

The resulting doping dependencies of \dea~and \deb~are shown in Fig. \ref{fig:diag}b and the values are reported in Table \ref{tab:tab1}. While both quantities show a maximum around $p\sim 0.13$, a doping-dependent anisotropy is observed. To make it clear, the thermodynamic anisotropy $N=~$\deb$-$\dea~is plotted in Fig. \ref{fig:diag}c.
Upon decreasing the doping level, $N$ first decreases and features a minimum for $p\sim 0.14$.
Then $N$ rises and show a maximum at $p\sim 0.11$, where \deb~is at least an order of magnitude larger than $|$\dea$|~\leq50$ K. Finally for $p<0.11$, $N$ decreases steadily as a mean-field jump is no longer resolved neither in $c_{11}$ nor in $c_{22}$ at $p=0.071$.
Thus, $N(p)$ is non-monotonic as a function of doping, which is the main experimental result of this article.

The behavior of $N(p)$ is to be contrasted with the monotonic increase of the 
orthorhombicity of YBCO with doping over similar range (see \cite{jorgensen90,casalta96,kruger97} and Fig. 4 in \cite{SM}). 
This difference in the doping trends imply that $N(p)$ is affected by an electronic property 
which we try to identify in the rest of
the paper. Below we discuss three possible electronic scenarios.

One possible source of additional electronic anisotropy can be the short range charge density wave (CDW) order
in YBCO \cite{ghinringhelli12,chang12,Wu15}. 
At face value this seems to be the case since \deb~ and \dea~ are individually peaked around $p =0.13$, which coincides
with the peak in the CDW ordering temperature. However, this simply implies that the 
CDW contributes significantly in the 
symmetric channel \deb$+$\dea, which is likely due to a competition between CDW and 
superconductivity \cite{cyrchoi18,vinograd19,kim18,kim21}. 
But, in the asymmetric channel \deb$-$\dea~ we do not expect the CDW to be 
important for the following reason.
The CDW state itself is either a biaxial order that preserves tetragonal symmetry \cite{forgan15}, in which 
case it does not contribute to $N(p)$,
or it is locally uniaxial with CDW domains running along the in-plane crystallographic axes as seen by 
X-ray  \cite{comin15,kim21}. However, even for the latter, the CDW will contribute to $N(p)$ only if these domains are aligned along the same direction, which is not the case in the zero strain limit probed here.
 
 A second possible explanation could be that the system is near a second order 
 electronic nematic phase transition, and that $N(p)$ is proportional to the associated order parameter
 that presumably increases as the doping level $p$ is reduced. In this scenario the system would have
 large  nematic correlation length in the $(x^2-y^2)$ symmetry channel.
 However, in this case one would expect the orthorhombic elastic constant to soften, 
 as seen in the iron based systems~\cite{yoshizawa12,boehmer14,gallais16}. No such 
 softening has been reported till date for any cuprate, while the absence of such softening is 
 well-established for La$_{2-x}$Sr$_x$CuO$_4$~\cite{nohara95,frachet21}.  Moreover, electronic Raman
scattering, which is a direct probe of nematicity~\cite{gallais16}, has shown the absence of nematic correlations in
underdoped Bi$_2$Sr$_2$CaCu$_2$O$_{8+\delta}$~\cite{auvray19}.
 Consistently, dynamical mean field studies have reported lack of any significant nematic
 correlations~\cite{gull09}, and the absence of nematic instability~\cite{okamoto10}.
 Consequently, while $N(p)$ is indeed an electronic anisotropy, it is unlikely to be due to the presence of a
 primary electronic nematic order parameter associated with a second order phase transition, 
 and in this sense the system is non-nematic.

The third possibility, which we explore in detail, is that $N(p)$ is governed by the opening of the pseudogap in the single particle electronic properties. 
This is based on the 
hypothesis that the pseudogap potential varies with external orthorhombic strain. 
With such an assumption we expect that
varying the pseudogap strength with orthorhombic strain will also change $T_c$, and this 
process will contribute to $N(p)$.
Qualitatively, in this scenario we expect that at low doping $N(p)$ vanishes with orthorhombicity 
for reasons of symmetry,
while at high doping $N(p)$ decreases because the pseudogap strength itself reduces with 
doping \cite{tallon01}. Thus,
$N(p)$ is guaranteed to have an extrema at intermediate doping. Quantitatively, our theory modeling 
of $N(p)$
consists of the following three steps.

First, we consider the free energy involving the superconducting order
parameter $\phi$ and the in-plane uniform strains $(u_{11}, u_{22})$.
To simplify the discussion we first assume a system with tetragonal
symmetry. The free energy has the form
\begin{align}
\label{eq:theory-1}
F =& \frac{1}{2} a \phi^2 + \frac{1}{2} c_{11} u_{11}^2 +
\frac{1}{2} c_{22} u_{22}^2 + c_{12} u_{11} u_{22}
\nonumber \\
+& \lambda_1 (u_{11} + u_{22} ) \phi^2 +
\frac{1}{2} \lambda_2 (u_{11} - u_{22})^2 \phi^2 + \cdots,
\end{align}
where the ellipsis implies terms irrelevant for the current discussion. Here
$a = a_0 (T - T_c^0)$, where $T_c^0$ is the superconducting transition
temperature in the absence of strain, $c_{11} = c_{22}$ and $c_{12}$
are elastic constants in Voigt notation, and $(\lambda_1, \lambda_2)$
are coupling constants. In an orthorhombic system we have
$u_{11} = u_0/2 + \epsilon_{11}$, and $u_{22} = - u_0/2 + \epsilon_{22}$,
where $u_0$ is the spontaneous orthorhombic strain, and
$(\epsilon_{11}, \epsilon_{22})$ are strains that may
develop in response to external stresses. Thus, to linear order in the
induced strains $\epsilon_{ii}$ the transition temperature is
\[
T_c (\epsilon_{ii}) = T_c^0 - \frac{2 \lambda_1}{a_0}(\epsilon_{11}
+ \epsilon_{22}) - \frac{2 \lambda_2}{a_0} u_0 (\epsilon_{11} - \epsilon_{22}),
\]
and from which we obtain
\begin{equation}
\label{eq:theory-2}
N = 4 u_0 \lambda_2/a_0.
\end{equation}

Second, we deduce a microscopic expression for the parameter $a_0$.
Since the superconducting transition is an instability in the
particle-particle channel, we can write
\begin{equation}
\label{eq:theory-3}
a = 1/g - \frac{1}{k_B T} \sum_{{\bf k}, \omega_n} f_{\bf k}^2
G_{\bf k}(i \omega_n) G_{-\bf k}(-i \omega_n),
\end{equation}
where $g$ is the pairing potential, $k_B$ is Boltzmann constant, $f_{\bf k}$ is a $d$-wave form factor,
and $G_{\bf k}(i \omega_n)$ is the electron Green's function.
We use the Yang-Zhang-Rice model~\cite{yang06} type of model for the Green's function
\begin{equation}
\label{eq:theory-4}
 G_{\bf k}^R(\omega)^{-1} = \omega + i\Gamma_1 - \epsilon_{\bf k}
 - \frac{P_{\bf k}^2}{\omega + i \Gamma_2 +  \xi_{\bf k}},
 \end{equation}
which has been widely used in the literature to study the low-energy properties of the 
 pseudogap~\cite{kyung06,stanescu06,liebsch09,sakai09,sakai15,wu18,norman98,norman07}.
Here, $\epsilon_{\bf k}$ is the electron dispersion,
$\xi_{\bf k}= -\omega$ defines the
line along which the electron spectral function is suppressed at a
given frequency,
$(\Gamma_1, \Gamma_2)$ are inverse lifetimes, and the pseudogap potential
$P_{\bf k} \equiv f_{\bf k} P_0$ is assumed to have $d$-wave symmetry. Once the
Green's function is known, the quantity $a_0$ follows simply from
\begin{equation}
\label{eq:theory-5}
a_0 = (\partial a/\partial T)_{T=T_c^0}.
\end{equation}

Third, we obtain a similar microscopic expression for the parameter $\lambda_2$.
We consider a tetragonal system with an externally imposed orthorhombic strain
$\eta \equiv u_{11} - u_{22}$. For finite $\eta$ one expects mixing between $A_{1g}$
and $B_{1g}$ symmetries. Thus, the four-fold symmetric functions
$(\epsilon_{\bf k}, \xi_{\bf k})$ develop a $d$-wave component, while the pseudogap
potential $P_{\bf k}$ develops an $s$-wave component. We express these changes as
$\epsilon_{\bf k} \rightarrow \tilde{\epsilon}_{\bf k} = \epsilon_{\bf k} + \alpha_1 \eta f_{\bf k}$,
$\xi_{\bf k} \rightarrow \tilde{\xi}_{\bf k} = \xi_{\bf k} + \alpha_2 \eta f_{\bf k}$, and
$P_{\bf k} \rightarrow \tilde{P}_{\bf k} = P_{\bf k} + \beta \eta P_0$, where
$(\alpha_1, \alpha_2)$ are constant energy scales and $\beta$ is 
an important dimensionless constant capturing the change of pseudogap with \emph{external}
orthorhombic strain.
From Eq.~(\ref{eq:theory-1}) we get
\begin{equation}
\label{eq:theory-6}
\lambda_2 =  (1/2)(\partial^2 a/\partial D^2),
\end{equation}
where the derivative
\[
\frac{\partial}{\partial D} \equiv \alpha_1 f_{\bf k} \frac{\partial}{\partial \epsilon_{\bf k}}
+ \alpha_2 f_{\bf k} \frac{\partial}{\partial \xi_{\bf k}} + \beta P_0 \frac{\partial}{\partial P_{\bf k}}.
\]
Thus, Eqs.~(\ref{eq:theory-2})-(\ref{eq:theory-6}) and the experimental input of $u_0$ obtained from
diffraction data provide a means to compute the
thermodynamic anisotropy $N$. The details of the particular microscopic model used and the technical
steps for the computation of $a_0$ and $\lambda_2$ can be found in \cite{SM}.

\begin{figure}
\center
    \includegraphics[width=8cm]{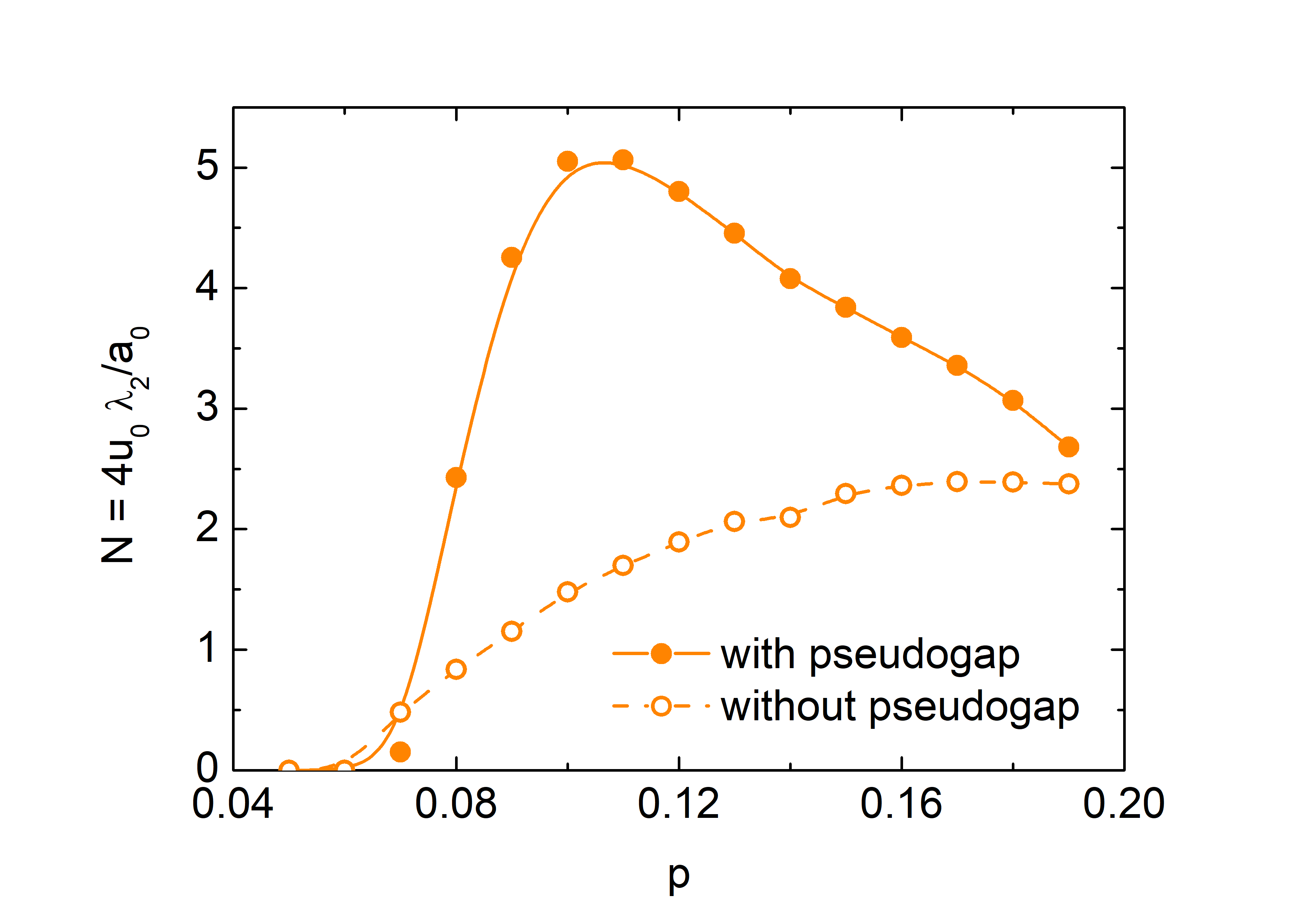}
    \caption{Theoretical $N=4u_0\lambda_2/a_0$ computed with $P_g \neq 0$ (full circles) and $P_g=0$ (empty circles), using a doping dependent orthorhombicity $u_0$ from scattering measurements \cite{SM}, and the pseudogap potential from \cite{tallon01}. Without pseudogap, $N$ increases monotonically, mimicking the doping dependent orthorhombicity. The effect of the pseudogap is to produce a non-monotonic $N$.}
    \label{fig:theory}
\end{figure}

Following our earlier hypothesis, we chose the constants $(\alpha_1, \alpha_2, \beta)$ such that the derivative
above is dominated by the last term which is the main pseudogap contribution. This would imply that the main
contribution to $N(p)$ can be captured by
\begin{equation}
\label{eq:theory-6a}
N(p) \approx (2/a_0)(\partial^2 a/\partial P_{\bf k}^2) \beta^2 u_0(p) P_0(p)^2.
\end{equation}
The results of the calculation are shown in Fig.~\ref{fig:theory}.  Our main
theoretical conclusion is that, in the presence of the pseudogap, the thermodynamic anisotropy $N(p)$ (the solid line)
has a maximum around $p=0.11$ doping, as seen in the experiments.
Beyond this doping the thermodynamic anisotropy decreases even though the crystalline anisotropy, namely the spontaneous orthorhombicity $u_0(p)$, increases until around $p=0.16$. 
The non-monotonic behavior of $N(p)$ is a result of the presence of the pseudogap. 
This point is clearly demonstrated by the monotonic evolution of the open symbols 
in Fig.~\ref{fig:theory} which are obtained by setting the pseudogap to zero. 
In other words, the doping dependence of $N(p)$ is controlled by that of the lattice
orthorhombicity $u_0(p)$ and the pseudogap potential $P_0(p)$, as expressed in Eq.~(\ref{eq:theory-6a}).
Thus, in Fig.~\ref{fig:theory} the initial increase of $N(p)$ for
$0.065 \leq p \leq 0.11$
is driven by the increase in the 
orthorhombicity $u_0(p)$, with the magnitude of $N(p)$ boosted by the presence of the pseudogap.
While, the later decrease of $N(p)$ (the solid line) with doping beyond $p=0.11$ is driven 
by a decrease of the pseudogap
potential $P_0$ and, therefore, a decrease of $\lambda_2(p)$.
The role of the pseudogap to enhance the in-plane electronic anisotropy 
has been also noted in an earlier dynamical mean field study~\cite{okamoto10}.

In the actual experiments $N(p)$ has a minimum around $p \sim 0.14$, and it increases
with further hole doping, a behavior reminiscent of electrical resistivity \cite{ando02} and thermal expansion \cite{nagelthesis}. 
In this regime the pseudogap decreases (see Fig. \ref{fig:diag}) and our model loses significance. Simultaneously,
the impact of the CuO chains, whose oxygen content increases with doping, becomes increasingly significant for the anisotropy.
A second possibility is that, with increasing doping the nematic correlations become stronger~\cite{auvray19}.
In addition to producing orthorhombicity, the CuO chains of YBCO also go through several structural changes in the range of oxygen content $y$ studied here \cite{zimmermann03}. The Ortho-II phase is found up to $y=6.6$ ($p<0.11$ or so). Then, increasing $y$ from 6.6 to 7.0, four other CuO chain superstructures are stabilized \cite{zimmermann03}. If this sequence of CuO chain structures had an impact on the thermodynamic anisotropy $N$, we would expect each one of them to cause an abrupt feature in $N$.  Instead, we observe a smooth evolution with a single broad minimum at $p  \sim 0.14$. Consequently, it is unlikely that $N(p)$ is affected by these structural changes in the chains.

To conclude, using ultrasounds on YBCO we extract $dT_{\rm c}/d \epsilon_{ii}$,
the variation of the superconducting transition temperature \Tc~with in-plane strain $\epsilon_{ii}$. We show that the in-plane thermodynamic anisotropy
$N \equiv dT_{\rm c}/d \epsilon_{22} - dT_{\rm c}/d \epsilon_{11}$ has an intriguing doping $p$
dependence that does not follow that of the crystalline orthorhombicity. 
Theoretically, we show that the data is qualitatively consistent with  Eq.~(\ref{eq:theory-6a}) 
which suggests that uniaxial strain affects the pseudogap which, in turn, affects $T_c$.

Finally, an important \emph{prediction} of our work is that, in the presence of  substantial uniaxial strain,
the pseudogap potential would vary significantly and, in particular, can lead to visible gap opening
in the nodal region. 
This prediction can be tested by performing 
angle resolved photoemission, electronic Raman response, in-plane resistivity 
and Hall measurements under uniaxial strain. 
Validation of the prediction would imply that uniaxial pressure is an important tool
to control the pseudogap, which is otherwise well known to be insensitive to external perturbations
such as disorder, magnetic field, and hydrostatic pressure.

%
%
We thank C. Meingast, M. Civelli, M.-H. Julien, A. Sacuto and Y. Gallais for valuable discussions. Part of this work was performed at the LNCMI, a member of the European Magnetic Field Laboratory (EMFL). Work at the LNCMI was supported by the Laboratoire d’Excellence LANEF (ANR-10-LABX-51-01), French Agence Nationale de la Recherche (ANR) grant ANR-19-CE30-0019-01 (Neptun) and EUR grant NanoX nANR-17-EURE-0009. Self-flux growth was performed at Scientific facility crystal growth in
  Max Planck Institute for Solid State Research, Stuttgart, Germany with
the support of the technical staffs.

%
%



\clearpage
\newpage

\onecolumngrid
\appendix

\renewcommand{\figurename}{{\bf Supplementary Figure}}
\renewcommand{\thefigure}{S\arabic{figure}}
\setcounter{figure}{0}

\renewcommand{\tablename}{{\bf Supplementary Table}}
\renewcommand{\thetable}{S\arabic{table}}
\setcounter{table}{0}

\renewcommand\theequation{S\arabic{equation}}
\setcounter{equation}{0}

\setcounter{page}{1}
\section*{Supplementary material for}
\begin{centerline}
{\bf \large
Effect of pseudogap on electronic anisotropy
in the strain dependence}
  \end{centerline}
 
 \begin{centerline}
 {\bf \large  of the superconducting $T_c$ of underdoped YBa$_2$Cu$_3$O$_y$}
    \end{centerline}

\vskip0.5cm

\begin{centerline}
{M. Frachet$^1$ \emph{et al.,}}
\end{centerline}

$^1$LNCMI-EMFL, CNRS UPR3228, Univ. Grenoble Alpes, Univ. Toulouse, Univ. Toulouse 3, INSA-T,  Grenoble and Toulouse, France

\subsection{Measurement of $dT_{\rm c}/d \epsilon_{ii}$ without applying
static, uniform strain}
The method to extract $dT_{\rm c}/d \epsilon_{ii}$, the variation of superconducting transition temperature 
$T_c$ with uniform strain $\epsilon_{ii}$, $i= (1, 2, 3)$, without actually applying such strain is based on the 
Ehrenfest relation. This relation follows from thermodynamics of a second order superconducting  transition as 
we discuss below. Following Landau's theory of phase transitions the difference between the free energies of a
superconducting phase and a normal phase is
\begin{equation}
\label{Eq:A1}
F_S - F_N = \frac{1}{2} a_0 \left( T - T_c(\epsilon_{ii}) \right) \phi^2+ \frac{1}{4} b \phi^4 + \cdots,
\end{equation}
where $\phi$ is the superconducting order parameter, $(a_0, b)$ are constants, and the ellipsis denote terms of order 
$\phi^6$ and higher. The strain dependence of the transition temperature to linear order can be expressed as
\begin{equation}
\label{Eq:A2}
T_c(\epsilon_{ii}) = T_c(0) + \Gamma_1 \epsilon_{11} + \Gamma_2 \epsilon_{22} + \Gamma_3 \epsilon_{33},
\end{equation}
where $\Gamma_i$, $i= (1, 2, 3)$ are constants. Note, by symmetry the transverse strains do not enter at linear order
provided the superconducting order parameter is one-component, which is the case for an orthorhombic system.

Minimizing the free energy difference $(F_S - F_N)$ with respect to $\phi$ yields the standard answer
$\phi^2 =0$, for  $T \geq T_c$, and $\phi^2 = - a_0 \left( T - T_c(\epsilon_{ii}) \right)/b$ for $T \leq T_c$.
Thus, for $T \leq T_c$ we have 
\begin{equation}
\label{Eq:A3}
F_S - F_N = - \frac{\left[ a_0 \left( T - T_c(\epsilon_{ii}) \right) \right]^2}{4b} + \cdots,
\end{equation}
where the ellipsis denote terms of order $(T - T_c)^3$ and higher. 

We recall that the elastic constant $c_{ii}$
associated with the strain $\epsilon_{ii}$ is defined by $c_{ii} = (1/V_m) (\partial^2 F/ \partial \epsilon_{ii}^2)_{ \epsilon_{ii}=0}$, 
where $V_m$ is the molar volume. Thus, from Eq.~(\ref{Eq:A3}) we get that the jump in the elastic constant at $T_c$ is
\begin{equation}
\label{Eq:A4}
\Delta c_{ii}(T_c) \equiv \left( c_{ii} \right)_S - \left( c_{ii} \right)_N 
= \frac{1}{V_m} \frac{\partial^2 (F_S - F_N)}{\partial \epsilon_{ii}^2}
= - \frac{a_0^2}{2b V_m} \left( \frac{\partial T_c}{\partial \epsilon_{ii}} \right)^2 .
\end{equation}
In the above $\left( c_{ii} \right)_S \equiv c_{ii}(T_c^-)$ and $\left( c_{ii} \right)_N \equiv c_{ii}(T_c^+)$, where 
$T_c^-$ and $T_c^+$ are temperatures infinitesimally below and above $T_c$, respectively. Note, since the jump is defined 
at $T=T_c$, the terms in $(F_S - F_N)$ of order $(T - T_c)^3$ and higher drop out. Likewise, variation of $T_c$ at quadratic
and higher orders in strain do not contribute to the jump either. Finally, from the definition of specific heat we get that the 
mean field jump of the specific heat at constant pressure at the transition is
\begin{equation}
\label{Eq:A5}
\Delta C_P \equiv C_P^S - C_P^N = - T_c \frac{\partial^2}{\partial T^2} (F_S - F_N) = \frac{a_0^2 T_c}{2b}.
\end{equation}
Using Eqs.~(\ref{Eq:A4}) and (\ref{Eq:A5}) we get the Ehrenfest relation
\begin{equation}
\label{Eq:A6}
\Delta c_{ii}(T_c) = - \frac{\Delta C_P}{V_m T_c} \left( \frac{\partial T_c}{\partial \epsilon_{ii}} \right)^2.
\end{equation}
The above equation implies that if the jump in the elastic constant $\Delta c_{ii}$ and that in the specific heat $\Delta C_P$
are known, then the magnitude of $\partial T_c / \partial \epsilon_{ii}$ can be deduced. $\Delta C_P$ is obtained from 
standard specific heat measurement. While the elastic constants, as a function of temperature, are measured using ultrasound 
technique. This involves creating acoustic waves which follow the dispersion relation $\omega = vq$,
and measuring their velocities $v = \sqrt{c/\rho}$, where $c$ is a suitable combination of the elastic constants, and $\rho$ is the 
density. Since the acoustic waves are nothing but strain waves at finite frequency and wavevector, the measurement does not 
involve applying static, uniform strain. Finally, the absolute sign of $\partial T_c / \partial \epsilon_{ii}$ is fixed by comparing
with existing data of how $T_c$ varies with uniaxial strain. Thus, we can extract $dT_{\rm c}/d \epsilon_{ii}$ without actually 
applying finite $\epsilon_{ii}$. The advantage of our method, compared to measuring variation of $T_c$
by applying $\epsilon_{ii}$, is that in our case the response is free from nonlinear effects of a finite $\epsilon_{ii}$, 
while in the latter it is
well-known that the response is dominated by non-linear effects which are not easy to interpret.

\subsection{Elastic constant anomaly at \Tc~for $p=0.185$ and $p=0.095$}

\begin{figure}[h]
\center
    \includegraphics[width=8cm]{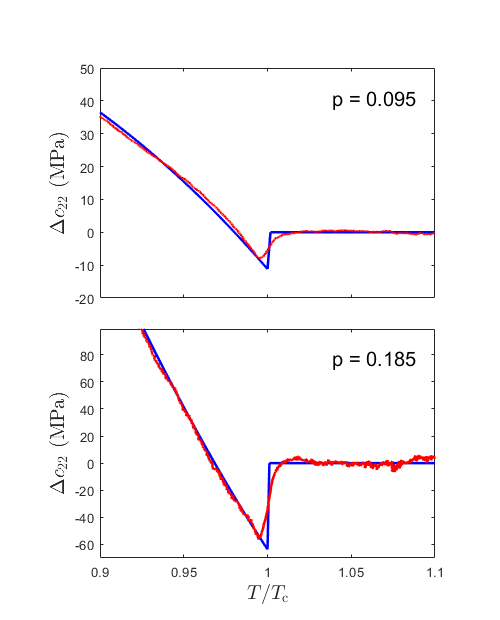}
    \caption{Superconducting contribution to $c_{22}$ for YBCO $p=0.095$ (top panel) and $p=0.185$ (bottom panel). The thermodynamic fit using Eq. 3 is shown in blue, data are shown in red.}
    \label{fig:6p4}
\end{figure}

\subsection{Experimental details}

The samples used in this study are detwinned single crystals of YBa$_2$Cu$_3$O$_y$ grown from high-purity starting materials. Note that for each oxygen concentration several samples with slightly different \Tc~and doping were measured. In total 20 samples were studied.

Sound velocity variation $\Delta v_s/v_s$ was measured using a standard pulse-echo technique \cite{luthi}. For high symmetry modes the sound velocity $v_s$ and the elastic constant $c_{ii}$ are related according to $\Delta v_s/v_s= \Delta c_{ii}/2c_{ii}$.  We focused on the sound velocity of the longitudinal mode propagating along the $b$-axis ($c_{22}$) and along the $a$-axis ($c_{11}$). The strain is defined as $\epsilon_{ii}=\frac{l_0 - l}{l_0}$ with $l_0$ the initial lattice parameter. $\epsilon_{ii} < 0$ indicates a tensile strain while $\epsilon_{ii}>0$ indicates a compressive strain.

 \subsection{Thermodynamic model used to fit the data}

 In order to extract \dei, we estimate the magnitude of $\Delta c_{ii}($\Tc) using an idealized mean-field second order jump fit to the data as done previously \cite{nohara95,kraut93}. The thermodynamic fit is derived from a two-fluid model of the free energy, in which \Tc~and the condensation energy $\phi$ are functions of the strain $\epsilon_{\rm ii}$:

\begin{equation}
\Delta F=-\phi(\epsilon_{\rm ii})\times(1-T^2/T^2_{\rm c}(\epsilon_{\rm ii}))^2
\end{equation}

The elastic constant is obtained by calculating the second derivative of the free energy with respect to $\epsilon_{\rm ii}$:

\begin{gather}
\Delta c_{ii}(T) = -\bigg (\frac{d \textrm{ln} T_{\rm c}}{d\epsilon_{\rm ii}}\bigg)^2\frac{T\Delta C_p(T)}{V_{\rm mol}}\notag \\
+ AT\Delta S(T) + \frac{1}{\phi} \frac{d^2\phi}{d\epsilon_{\rm ii}^2}\Delta F(T)
\end{gather} where $A$ is proportional to the strain derivatives of \Tc~and $\phi$ \cite{nohara95}. At $T=$ \Tc~this equation is equivalent to the Ehrenfest relationship.

\subsection{Error bars}
\begin{figure}[h]
\center
    \includegraphics[width=8cm]{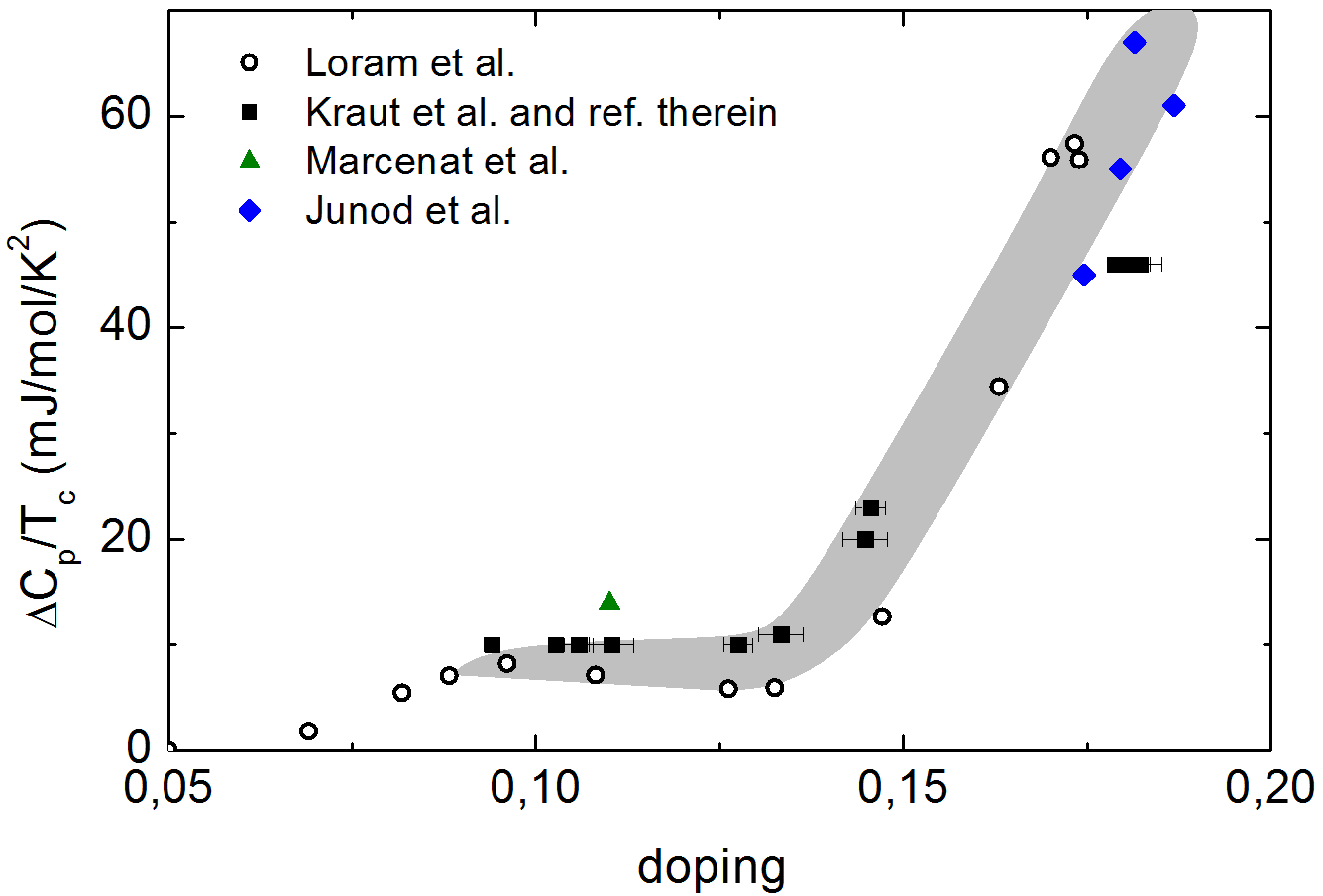}
    \caption{Reproduction of specific heat jump at \Tc~($\Delta$C$_p$/\Tc) as a function of doping from the literature\cite{junod89,wuhl91,claus92,loram01,marcenat15}. The error bars on $\Delta$C$_p$/\Tc~are evaluated using the scattering of the different data set, highlighted by the gray shaded area.}
    \label{fig:Cp}
\end{figure}

The error bars on \dei~are estimated as follows. Most of the uncertainty comes from the value of the specific heat jump at \Tc, $\Delta C_p(T_{\rm c})$. The specific heat was not measured in the samples used for this study. We relied on specific heat data from the literature. $\Delta C_p(T_{\rm c})/T_{\rm c}$ are shown in figure \ref{fig:Cp} where the shaded area highlights the scattering of the data, and is used to estimate the error bar on $\Delta C_p(T_{\rm c})/T_{\rm c}$.

Another source of error comes from the uncertainty on the absolute value of the elastic constants $c_{11}$ and $c_{22}$. The pulse-echo technique used in this study allows to measure the absolute value of the sound velocity with an accuracy of a few $\%$. This originates from the uncertainty on the sample dimension and from the fact that we used transducers with finite thickness, resulting in irregular echo shape. Comparing our data in YBCO 6.99 with data of Lei \emph{et al.} \cite{lei93} and ref. therein, we estimate an error $\Delta c_{\rm ii}/c_{\rm ii}\approx 6~\%$. We took into account the doping dependence of $c_{11}$ and $c_{22}$ using a parabolic model \cite{nagelthesis}

 \begin{equation}
 c_{ii}(y)= \langle c \rangle + (y-6)^2\big( c_{ii}(y=7)-\langle c \rangle \big)
 \end{equation} with
 \begin{equation}
  \langle c \rangle = \frac{ c_{11}(y=7)+ c_{22}(y=7)}{2}
 \end{equation}
 $y$ is the oxygen content in YBa$_{2}$Cu$_{3}$O$_{y}$, and we used $c_{11}(y=7)=215$~GPa and $c_{22}(y=7)=255$~GPa. The previous formula reflects the doping dependence of $c_{ii}$ due to the orthorhombicity of YBCO. $c_{11}$ and $c_{22}$ must converge to the same value at low doping level, and are increasingly different with increasing doping. This formula results in $6~\%$ change in $c_{\rm ii}$ across the doping range studied here. This doping dependence was not observed experimentally most likely because of the low accuracy of the pulse-echo method. Taken into account this doping dependence has little effect on the resulting doping dependence of \dei, given the large doping dependence of the latter. Nonetheless, we took it into account for the sake of completeness of the analysis.

Other sources of error include experimental errors (variations of the amplitude of the elastic constant jump at \Tc~for different samples at similar doping level), uncertainties on the evaluation of the thermal phonon background which is subtracted to the data to isolate the superconducting contribution, and errors from the thermodynamic fit.
\subsection{Comparison with uniaxial pressure results} 
\begin{figure}[h]
\center
    \includegraphics[width=8cm]{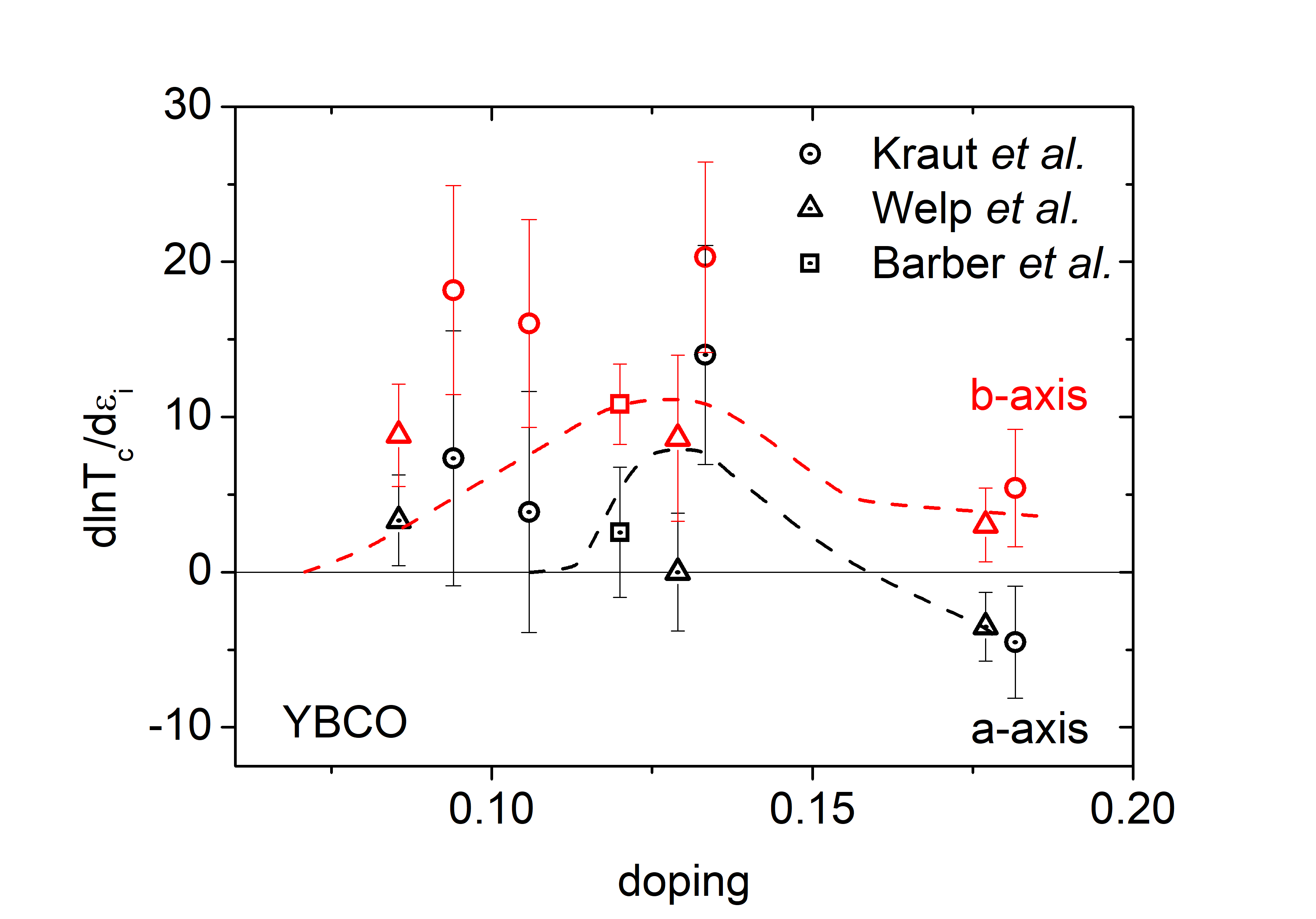}
    \caption{Data from uniaxial pressure (triangles \cite{welp94}, squares \cite{barber21}) and thermal expansion (circles) \cite{kraut93} measurements converted in uniaxial strain dependences. Red is for \deb~and black (with center dot) for \dea. The error bars are large due to the uncertainty on the elastic tensor and on the d\Tc$/$d$P_i$. Dashed lines are the same as those shown in Fig. 3 of the main text.}
    \label{fig:dei}
\end{figure}

The derivatives \dei~and d\Tc$/$d$P_i$ are related via the formula \dei$=\sum_j c_{ij}$d\Tc$/$d$P_j$. Consequently, in order to compare our results with those from uniaxial pressure measurements we need the complete elastic tensor of YBCO. For the calculation of  d\Tc$/$d$P_i$ we use data from Lei \emph{et al.} \cite{lei93} and ref. therein, obtained in overdoped YBCO, and we assume doping independent elastic constants. The uncertainties on the off-diagonal elastic constants are large and result in large error bars in the \dei~obtained this way. In Fig.\ref{fig:dei}, we plot \dei~estimated from measurements of thermal expansion \cite{meingast91,kraut93}  and direct measurements under uniaxial pressure \cite{welp92,welp94,barber21}. Values are also reported in Table \ref{tab:tab2}. There is an overall agreement between all the data sets.

\begin{table}[h]
    \begin{center}
        \begin{tabular}{cccc}
\hline

$~~~p$~(holes/Cu) & \dea (K) & \deb (K) & Ref.\\

\hline
\hline

0.086 & $170 \pm 149$ & $450 \pm 168$ & \cite{welp94}\\
0.094  & $407 \pm 455$ & $1008 \pm 374$ & \cite{kraut93}\\
0.106  & $233 \pm 465$ & $961 \pm 401$ & \cite{kraut93}\\
0.120 & $168 \pm 275$ & $709 \pm 170$ & \cite{barber21}\\
0.129 & $0 \pm 275$ & $626 \pm 388$ & \cite{welp94}\\
0.133  & $1068 \pm 537$ & $1548 \pm 469$ & \cite{kraut93}\\
0.177 & $-320 \pm 201$ & $277 \pm 217$ & \cite{welp94}\\
0.182  & $-405 \pm 324$ & $486 \pm 340$ & \cite{kraut93}\\

\hline
\hline

    \end{tabular}
    \end{center}
\caption{\dei~calculated from $\sum_j c_{ij}$d\Tc$/$d$P_j$ where d\Tc$/$d$P_i$ was measured in direct uniaxial pressure experiments \cite{welp94,barber21} or with thermal expansion \cite{kraut93}. Elastic constants value taken from \cite{lei93}.}
\label{tab:tab2}
\end{table}

\subsection{Orthorhombicity}

\begin{figure}[h]
\center
    \includegraphics[width=8cm]{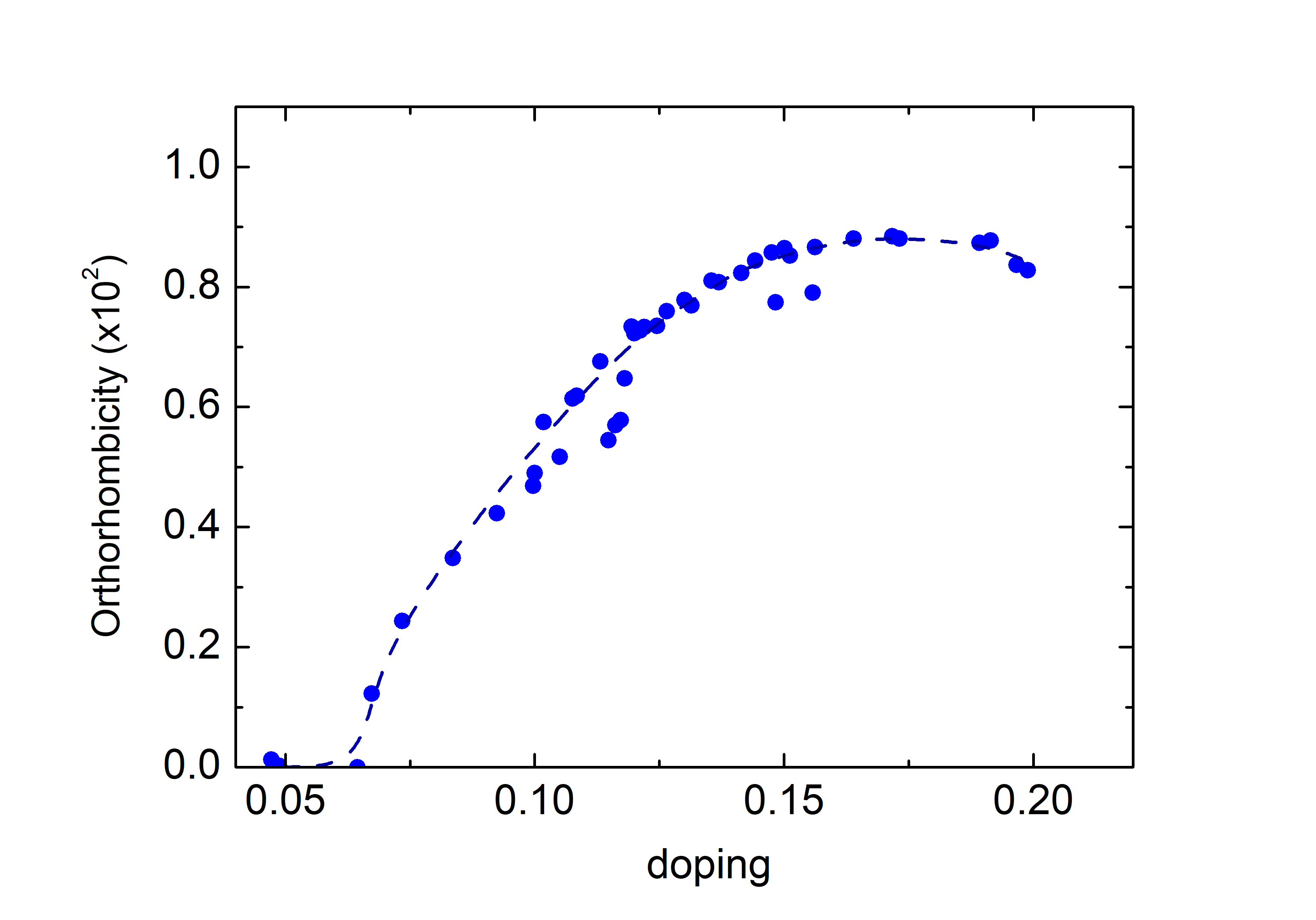}
    \caption{Blue circles show the orthorhombicity $u_0=(b-a)/(b+a)$ as a function of hole doping, measured in diffraction experiments \cite{jorgensen90, casalta96,kruger97}. The blue dotted line is a guide to the eye. This dotted line is used to calculate $N\propto u_0$.}
    \label{fig:u0}
\end{figure}

The orthorhombicity from diffraction experiments is shown in blue in Fig. \ref{fig:u0}. At low doping level, the orthorhombicity increases steadily as oxygen content in the CuO chains is increased. However, for doping levels $p>0.15$ or so, the orthorhombicity saturates whereas the oxygen content keeps increasing.This saturation can be caused by the pressure of the oxygen ordering process in the CuO chains of YBCO \cite{nagel00,nagelthesis}. Increasing oxygen content results in an increase in the anisotropy of in-plane electronic transport \cite{ando02} and in-plane thermal expansivities \cite{nagelthesis}, even in the doping range where the orthorhombicity saturated. Assuming the sound velocity has a similar doping-dependent anisotropy as thermal expansivity, the increase of the measured $N$ for $p>0.15$ can be naturally explained.

For practical purposes, the anisotropy from diffraction experiments \cite{jorgensen90, casalta96,kruger97} (blue dotted line in Fig. \ref{fig:u0})  is used for computing the theoretical doping dependence of $N$, such as shown in Fig. 4 of the main text.

\subsection{Theoretical details}
Using Landau-Ginzburg type argument we established that the measured thermodynamic anisotropy
$N \equiv dT_{\rm c}/d \epsilon_{22} - dT_{\rm c}/d \epsilon_{11}$ can be expressed as (see Eq.~(3) of main text)
\[
N = 4 u_0 \lambda_2/a_0.
\]
In the above $u_0$ is the orthorhombicity, which is known experimentally. Below
we provide the technical details for computing the quantities $a_0$ and $\lambda_2$ starting from a
microscopic model. These are defined by Eqs.~(6) and (7), respectively, of the main text.

Our starting point is the assumption that the low energy electrons can be described by the Green's function
(see Eq.~(5) of the main text)
\[
G_{\bf k}^R(\omega)^{-1} = \omega + i\Gamma_1 - \epsilon_{\bf k}
 - \frac{P_{\bf k}^2}{\omega + i \Gamma_2 +  \xi_{\bf k}}.
\]
This ansatz has been widely used in the literature to capture the low energy properties of the cuprates in the
pseudogap state. The above can be rewritten as
\begin{equation}
\label{th-eq-1}
G_{\bf k}^R(\omega) = \frac{A_{1{\bf k}}}{\omega - \omega_{1{\bf k}}}
+ \frac{A_{2{\bf k}}}{\omega - \omega_{2{\bf k}}},
\end{equation}
where
\begin{equation}
\label{th-eq-2}
\omega_{1{\bf k},2{\bf k}} = \frac{1}{2}
\left[(z_{1{\bf k}} + z_{2{\bf k}}) \pm \sqrt{(z_{1{\bf k}} - z_{2{\bf k}})^2 + 4 P_{\bf k}^2} \right],
\end{equation}
with $z_{1{\bf k}} \equiv \epsilon_{\bf k} - i\Gamma_1$, $z_{2{\bf k}} \equiv -\xi_{\bf k} - i\Gamma_2$,
and
\begin{equation}
\label{th-eq-3}
A_{1{\bf k}} = \frac{\omega_{1{\bf k}} - z_{1{\bf k}}}{\omega_{1{\bf k}} - \omega_{2{\bf k}}},
\quad
A_{2{\bf k}} = \frac{z_{2{\bf k}} - \omega_{2{\bf k}}}{\omega_{1{\bf k}} - \omega_{2{\bf k}}}.
\end{equation}
In terms of the Green's function the particle-particle susceptibility is given by
\begin{equation}
\label{th-eq-4}
\chi_{pp}[\epsilon_{\bf k}, \xi_{\bf k},  P_{\bf k}] =
\frac{2}{\beta V} \sum_{{\bf k}, \omega_n}^{|\epsilon_{\bf k}| \leq \Lambda}
f_{\bf k}^2 G_{\bf k}(i \omega_n) G_{-\bf k}(-i \omega_n),
\end{equation}
where $\beta$ is inverse temperature, $V$ is volume, and
we assume that the Cooper pairing potential is zero above a cutoff energy scale $\Lambda$.
The form factor $f_{\bf k} \equiv \cos(k_x) - \cos(k_y)$ implies that the pairing instability is
in the $d$-wave channel.
Note, $\chi_{pp}$ is a functional of the dispersions $(\epsilon_{\bf k}, \xi_{\bf k})$, and the
pseudogap function $P_{\bf k}(p) = P_0 (p) f_{\bf k}$. Here $P_0(p)$ is the energy scale of the pseudogap
potential that varies with hole doping.

First, we discuss the details of the computation of $\lambda_2$.
We assume that $T_c^0$, the superconducting transition temperature in the absence of external strains, is
the lowest energy scale in the problem. Then, for the computation of $\lambda_2$ it is sufficient to set
temperature $T=0$. In this limit the above frequency sum can be performed analytically, and we get
\begin{equation}
\label{th-eq-5}
\chi_{pp} = \frac{2}{\pi V} \sum_{\bf k}^{|\epsilon_{\bf k}| \leq \Lambda} f_{\bf k}^2
\left[ \left(\frac{A_{1{\bf k}} A_{1{\bf k}}^{\ast}}{E_{1{\bf k}}} + 2X_{\bf k}^{\prime} \right)\cot^{-1}
\left( \frac{\gamma_{1{\bf k}}}{E_{1{\bf k}}} \right) +
\left(\frac{A_{2{\bf k}} A_{2{\bf k}}^{\ast}}{E_{2{\bf k}}} + 2X_{\bf k}^{\prime} \right)\cot^{-1}
\left( \frac{\gamma_{2{\bf k}}}{E_{2{\bf k}}} \right)
- X_{\bf k}^{\prime \prime} \ln \left( \frac{E_{2{\bf k}}^2
+ \gamma_{2{\bf k}}^2}{E_{1{\bf k}}^2 + \gamma_{1{\bf k}}^2} \right) \right],
\end{equation}
where $E_{1{\bf k}/2{\bf k}}$ and $\gamma_{1{\bf k}/2{\bf k}}$ are real quantities that are defined by
$\omega_{1{\bf k}/2{\bf k}} \equiv E_{1{\bf k}/2{\bf k}} - i \gamma_{1{\bf k}/2{\bf k}}$, and the complex
quantity $X_{\bf k} = X_{\bf k}^{\prime} + i X_{\bf k}^{\prime \prime}
\equiv A_{1{\bf k}}A_{2{\bf k}}/[E_{1{\bf k}} + E_{2{\bf k}} + i (\gamma_{2{\bf k}} - \gamma_{1{\bf k}})]$.
In the presence of a finite external orthorhombic strain $\eta$ the quantities
$(\epsilon_{\bf k}, \xi_{\bf k},  P_{\bf k})$ transform to
$(\tilde{\epsilon}_{\bf k}, \tilde{\xi}_{\bf k},  \tilde{P}_{\bf k})$, where
$\tilde{\epsilon}_{\bf k} = \epsilon_{\bf k} + \alpha_1 \eta f_{\bf k}$,
$\tilde{\xi}_{\bf k} = \xi_{\bf k} + \alpha_2 \eta f_{\bf k}$, and
$\tilde{P}_{\bf k} = P_{\bf k} + \beta \eta P_0$. From Eq.~(7) of the main text we get
$\lambda_2 =  -1/2(\partial^2 \chi_{pp}/\partial^2 \eta)_{\eta=0}$. This  implies that
\begin{equation}
\label{th-eq-6}
\lambda_2 = - \frac{1}{\pi V} \sum_{\bf k}^{|\epsilon_{\bf k}| \leq \Lambda} f_{\bf k}^2
\left( \alpha_1 f_{\bf k} \frac{\partial}{\partial \epsilon_{\bf k}} +
\alpha_2 f_{\bf k} \frac{\partial}{\partial \xi_{\bf k}} + \beta P_0 \frac{\partial}{\partial P_{\bf k}}
\right)^2 L_{\bf k},
\end{equation}
where $L_{\bf k}$ denotes the quantity within $\left[ \cdots \right]$ in Eq.~(\ref{th-eq-5}). In the above
equation it is straightforward to take the derivatives and then perform the momentum sum numerically.
This leads to the evaluation of $\lambda_2(p)$ as a function of hole doping $p$.

Next, we discuss the details of the computation of $a_0 = - (\partial \chi_{pp}/\partial T)_{T=T_c^0}$.
In terms of the Green's function this can be written as
\begin{equation}
\label{th-eq-8}
a_0 = \frac{2}{V} \sum_{\bf k}^{|\epsilon_{\bf k}| \leq \Lambda} f_{\bf k}^2 \, {\rm Im}
\int_{-\infty}^{\infty} \frac{d \omega}{2\pi}  G_{\bf k}^R(\omega)  G_{\bf k}^A(-\omega)
\frac{\omega}{2T^2 \cosh^2 (\omega/(2T))}.
\end{equation}
The thermal factor ensures that the $\omega$-integral contributes only for $\omega \lesssim T = T_c^0$.
Since $T_c^0$ is the lowest energy scale, the Green's functions can be expanded in powers of the frequency. This is equivalent to an expansion in powers of $T_c^0/{\rm max}[\Gamma_1,P_0]$.
We keep the first non-zero term, and we get
\begin{equation}
\label{th-eq-9}
a_0 = \frac{2 T_c^0}{3\pi V} \sum_{\bf k}^{|\epsilon_{\bf k}| \leq \Lambda} f_{\bf k}^2
\left[ \frac{A_{1{\bf k}} A_{1{\bf k}}^{\ast} \gamma_{1{\bf k}}}
{(E_{1{\bf k}}^2 + \gamma_{1{\bf k}}^2)^2} +
\frac{A_{2{\bf k}} A_{2{\bf k}}^{\ast} \gamma_{2{\bf k}}}
{(E_{2{\bf k}}^2 + \gamma_{2{\bf k}}^2)^2}
+ {\rm Im} \left\{  \frac{A_{1{\bf k}} A_{2{\bf k}}^{\ast}}
{(E_{1{\bf k}} - i \gamma_{1{\bf k}}) (E_{2{\bf k}} + i \gamma_{2{\bf k}})}
\left(
\frac{1}{E_{1{\bf k}} - i \gamma_{1{\bf k}}}
- \frac{1}{E_{2{\bf k}} + i \gamma_{2{\bf k}}}
\right) \right\} \right].
\end{equation}
It is simple to perform the momentum sum numerically, which leads to $a_0(p)$ as a function of
hole doping $p$.

\begin{figure}
\center
\includegraphics[width=\textwidth]{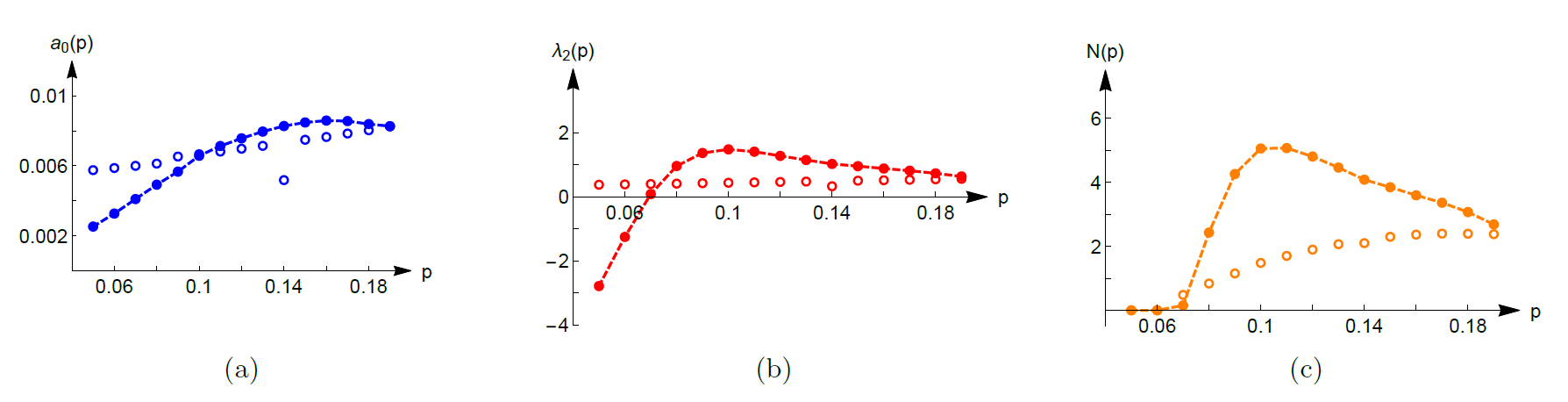}
\caption{\label{SI-fig4}
(a, b, c) Variations of the quantities $a_0(p)$, $\lambda_2(p)$ and the thermodynamic
anisotropy $N(p)$ as functions of hole doping,
respectively. Filled symbols are with finite pseudogap, and open symbols are calculations with
the pseudogap potential set to zero. The theoretical model correctly captures the appearance of a
maximum in $N(p)$, as seen experimentally. This feature disappears, and $N(p)$ is a monotonic
function of $p$ when the pseudogap is set to zero.}
\end{figure}

We compute $a_0$, $\lambda_2$ and the thermodynamic anisotropy $N(p)$
using the following model and parameters. The dispersions are taken as
\begin{subequations}
\begin{align}
\label{th-eq-10}
\epsilon_{\bf k} &= -2t (\cos (kx) + \cos (ky)) + 4 t^{\prime} \cos (kx) \cos (ky)
-2t^{\prime \prime} (\cos (2 kx) + \cos (2 ky)) - \mu,\\
\xi_{\bf k} &= -2t (\cos (kx) + \cos (ky)),
\end{align}
\end{subequations}
with $t=1$, $t^{\prime} = 0.3t$, $t^{\prime \prime} = 0.2t$. The damping factors are set to
$\Gamma_1 = 0.1 t$, and $\Gamma_2 = 0.01 t$. The pseudogap potential is set to
$P_0(p) = P_g (1 - p/0.2)$, with $P_g =0.3 t$. Thus, $P_0(p)$ is
assumed to decrease linearly with hole doping, and disappearing at $p=0.2$. Next, we
take the energy scales $\alpha_1 = \alpha_2 = 0.5 t$, and the dimensionless parameter
$\beta= -10$, and the overall energy
cutoff $\Lambda =0.3t$. For the computation of $N(p)$ we use the
experimental values of the spontaneous orthorhombicity $u_0(p)$ of YBa$_2$Cu$_3$O$_y$.

The results of the calculation are shown in Fig.~\ref{SI-fig4}. Our main conclusion is that in the
presence of the pseudogap the thermodynamic anisotropy $N(p)$
indeed has a maximum around $p=0.11$ doping, see the evolution of the filled symbols
in Fig.~\ref{SI-fig4}(c). Beyond this doping the thermodynamic anisotropy
decreases even though the crystalline anisotropy, namely the spontaneous orthorhombicity $u_0(p)$
increases until $p=0.18$. The non-monotonic behavior of $N(p)$ is a result of the presence of
the pseudogap. This point is clearly demonstrated by the monotonic evolution of the open symbols
in Fig.~\ref{SI-fig4}(c) which are obtained by setting the pseudogap to zero.

The decrease of $N(p)$ for $p>0.1$ in our calculation is the result of the following two features.
First, the increase of $a_0(p)$ in this doping range. This is due to the fact that the pseudogap decreases
with increasing doping and, therefore, there is more phase space for the contribution of the low energy
electrons to the susceptibility $\chi_{pp}$ and to its temperature dependence. In general, we expect that
susceptibilities are less temperature dependent in the presence of gaps. Second, the decrease in the
magnitude of $\lambda_2(p)$ over the same doping range. This feature is the result of our assumption
that the pseudogap potential varies significantly in the presence of an external uniaxial strain. Thus,
around $p \approx 0.11$ the contribution to $\lambda_2$ is dominated by the term
$(\partial/\partial P_{\bf k})^2$ in Eq.~(\ref{th-eq-6}) in our model. On the other hand, by definition, at
$p=0.2$ this contribution [and also from terms involving $(\partial/\partial \xi_{\bf k})$]
vanishes. In other words, an important \emph{prediction} of our work is that, in the presence of
substantial uniaxial strain the pseudogap potential would vary significantly and, in particular, can lead to
visible gap opening in the nodal region. This prediction can be tested by performing spectroscopy such as
angle resolved photoemission and electronic Raman response under uniaxial strain.

In the actual experiments $N(p)$ has a minimum around $p \sim 0.14$ and then increases with further
hole doping. We think this regime is dominated by the contribution of the anisotropy coming from the
CuO chains, rather than the electrons of the copper-oxygen planes. Consequently, this increase is not
captured in our theoretical modeling.


\begin{references}

\bibitem{ando02}
Y. Ando \emph{et al.} Phys. Rev. Lett. \textbf{88} 137005 (2002)

\bibitem{sato17}
Y. Sato, S. Kasahara, H. Murayama, Y. Kasahara, E.-G. Moon, T. Nishizaki, T. Loew, J. Porras,
B. Keimer, T. Shibauchi and Y. Matsuda,
Nat. Phys. {\bf 13}, 1074 (2017)

\bibitem{hinkov08}
V. Hinkov \emph{et al.}, Science \textbf{319}, 597 (2008)

\bibitem{Mangin-Thro17}
L. Mangin-Thro, Y. Li, Y. Sidis, and P. Bourges,
Phys. Rev. Lett. {\bf 118}, 097003 (2017)

\bibitem{Achkar16}
A. J. Achkar \emph{et al.}, Science \textbf{351} 576 (2016)

\bibitem{daou10}
R. Daou \emph{et al}., Nature \textbf{463} 519 (2010)

\bibitem{cyrchoi15}
O. Cyr-Choiniere \emph{et al.} Phys. Rev. B \textbf{92} 224502 (2015)

\bibitem{lawler10}
M. J. Lawler \emph{et al.}, Nature {\bf 466}, 347 (2010).

\bibitem{zheng17}
Y. Zheng, \emph{et al.}, Sci. Rep. {\bf 7}, 8059 (2017).

\bibitem{Wu15}
T. Wu, \emph{et al.}, Nat. Comm. {\bf 6}, 6438 (2015).

\bibitem{ishida20}
K. Ishida \emph{et al.,} JPSJ {\bf 89} 064707 (2020).

\bibitem{Wu17}
J. Wu \emph{et al.}, Nature \textbf{547} 432 (2017).

\bibitem{nie14}
L. Niea,, G. Tarjus, and S. A. Kivelson,
Proc. Nat. Acad. Sc. (USA) {\bf 111}, 7980 (2014).

\bibitem{tranquada15}
E. Fradkin, S. A. Kivelson, and J. M. Tranquada,
Rev. Mod. Phys. {\bf 87}, 457 (2015).

\bibitem{morice18}
C. Morice, D. Chakraborty, X. Montiel, C. P\'{e}pin,
J. Phys. Condens. Matter {\bf 30}, 295601 (2018).

\bibitem{sachdev19}
S. Sachdev, H. D. Scammell, M. S. Scheurer, and G. Tarnopolsky,
Phys. Rev. B {\bf 99}, 054516 (2019).

\bibitem{orth19}
P. P. Orth, B. Jeevanesan, R. M. Fernandes, and J. Schmalian,
npj Quantum Mater. {\bf 4}, 4 (2019).

\bibitem{gull09}
E. Gull, O. Parcollet, P. Werner, and A. J. Millis,
Phys. Rev. B {\bf 80}, 245102 (2009).

\bibitem{gallais13}
 Y. Gallais, R. M. Fernandes, I. Paul, L. Chauvi\`{e}re, Y. -X. Yang, M. -A. M\'{e}asson,
 M. Cazayous, A. Sacuto, D. Colson, and A. Forget,
 Phys. Rev. Lett. {\bf 111}, 267001 (2013).

\bibitem{auvray19}
N. Auvray, S. Benhabib, M. Cazayous, R. D. Zhong, J. Schneeloch, G. D. Gu, A. Forget,
D. Colson, I. Paul, A. Sacuto, and Y. Gallais,
Nat. Comm. {\bf 10}, 5209 (2019).

\bibitem{kraut93}
Kraut, O., Meingast, C., Brauchle, G., Claus, H., Erb, A., Müller-Vogt, G., and Wühl, H. Uniaxial pressure dependence of \Tc~of untwinned  YBa$_2$Cu$_3$O$_{\rm x}$ single crystals for x=6.5–7. Physica C: Superconductivity \textbf{205}, 139–146 (1993)

\bibitem{pasler00}
V. Pasler, PhD thesis, Karlsruhe university (2000)

\bibitem{jorgensen90}
J. D. Jörgensen \emph{et al.}, Phys. Rev. B 41, 1863 (1990)

\bibitem{casalta96}
H. Casalta \emph{et al.}, Physica C 258, 321 (1996)

\bibitem{kruger97}
Ch. Krüger \emph{et al.}, Journal of Solid State Chemistry 134, 356 (1997)

\bibitem{liang06}
Liang, R., Bonn, D. A. and Hardy, W. N. Evaluation of CuO$_2$ plane hole doping in  YBa$_2$Cu$_3$O$_{\rm 6+x}$ single crystals Physical Review B  \textbf{73} 180505 (2006)

\bibitem{SM} 
See Supplemental Material at [URL will be inserted by publisher] for experimental details, 
additional data, fitting model, error bar estimation , comparison with uniaxial pressure results, and theory details.

\bibitem{varshni70}
Varshni, Y. P. Temperature Dependence of the Elastic Constants Phys. Rev. B \textbf{2} 3952-3958 (1970)

\bibitem{testardi71}
L. R. Testardi, Phys. Rev. B {\bf 3}, 95 (1971).

\bibitem{rehwald73}
W. Rehwald, Adv. Phys. {\bf 22}, 721 (1973).

\bibitem{millis88}
Millis, A. J. and Rabe, K. M. Superconductivity and lattice distortions in high-$T_c$ superconductors Phys. Rev. B  \textbf{38}, 8908-8919 (1988)

\bibitem{luthi}
Lüthi, B. Physical Acoustics in the Solid State, Springer Series for Solid-State Sciences, Vol. 148 (Springer, Berlin, New York, 2005).

\bibitem{nohara95}
Nohara, M.; Suzuki, T.; Maeno, Y.; Fujita, T.; Tanaka, I. and Kojima, H. Unconventional lattice stiffening in superconducting LSCO single crystals Phys. Rev. B \textbf{52} 570-580 (1995)

\bibitem{junod89}
A. Junod Physica C \textbf{162-164} 482 (1989)

\bibitem{wuhl91}
H. Wuhl Physica C \textbf{185-189} 482 (1991)

\bibitem{claus92}
H. Claus PhysicaC \textbf{198} 42 (1992)

\bibitem{loram01}
Loram, J.W., Luo, J., Cooper, J.R., Liang, W.Y., and Tallon, J.L.  Evidence on the pseudogap and condensate from the electronic specific heat. Journal of Physics and Chemistry of Solids \textbf{62} 59–64 (2001)

\bibitem{marcenat15}
Marcenat, C. et al. Calorimetric determination of the magnetic phase diagram of underdoped Ortho-II YBCO single crystals. Nat. Commun.  \textbf{6} 7927 (2015)

\bibitem{meingast90}
C. Meingast Phys. Rev. B \textbf{41} 11299 (1990)

\bibitem{meingast91}
C. Meingast Phys. Rev Lett. \textbf{67} 1634 (1991)

\bibitem{welp92}
U. Welp Phys. Rev. Lett. \textbf{69} 2130 (1992)


\bibitem{welp94}
U. Welp Journal of Superconductivity \textbf{7} 159 (1994).

\bibitem{ludwig96}
H. A. Ludwig J. of Low Temp. Phys. \textbf{105} 1385 (1996).

\bibitem{barber21}
MarK E. Barber \emph{et al.} preprint at arXiv:2101.02923 (2021).

\bibitem{footnote} \dea~for $p=0.18$ (empty black square, panel b) is determined using measurements of \Tc~under uniaxial stress \cite{welp92}, converted into strain dependence of \Tc~(see \cite{SM}). We use this value to calculate $N$ for $p\approx0.18$ (empty orange circle, panel c)

\bibitem{ghinringhelli12}
Ghiringhelli, G., Tacon, M.L., Minola, M., Blanco-Canosa, S., Mazzoli, C., Brookes, N.B., Luca, G.M.D., Frano, A., Hawthorn, D.G., He, F., et al. (2012). Long-Range Incommensurate Charge Fluctuations in (Y,Nd)Ba$_2$Cu$_3$O$_{\rm 6+x}$. Science \textbf{337}, 821–825.

\bibitem{chang12}
Chang, J., Blackburn, E., Holmes, A.T., Christensen, N.B., Larsen, J., Mesot, J., Liang, R., Bonn, D.A., Hardy, W.N., Watenphul, A., \emph{et al.}  Direct observation of competition between superconductivity and charge density wave order in YBa$_2$Cu$_3$O$_{6.67}$. Nat Phys \textbf{8}, 871–876 (2012).




\bibitem{cyrchoi18}
O. Cyr-Choinière Phys. Rev. B 98, 064513 (2018)

\bibitem{vinograd19}
I. Vinograd \emph{et al.} Phys. Rev. B 100, 094502 (2019)

\bibitem{kim18}
H.-H. Kim \emph{et al.} Science \textbf{362} 1040 (2018)

\bibitem{kim21}
H.-H. Kim \emph{et al.}, Phys. Rev. Lett. \textbf{126} 037002 (2021)

\bibitem{forgan15}
Forgan, E. M. \emph{et al.} The microscopic structure of charge density waves in underdoped YBa$_2$Cu$_3$O$_{y}$ revealed by X-ray diffraction. Nat. Commun. \textbf{6} 10064 (2015)


\bibitem{comin15}
R. Comin \emph{et al.} Science (2015).

\bibitem{yoshizawa12}
 M. Yoshizawa and S. Simayi, Mod. Phys. Lett. {\bf 26},
1230011 (2012).

\bibitem{boehmer14}
A. E. B\"{o}hmer, P. Burger, F. Hardy, T. Wolf, P. Schweiss,
R. Fromknecht, M. Reinecker, W. Schranz, C. Meingast,
Phys. Rev. Lett. {\bf 112}, 047001 (2014).

\bibitem{gallais16}
Y. Gallais and I. Paul, C. R. Phys. {\bf 17}, 113 (2016).

\bibitem{frachet21}
M. Frachet, S. Benhabib, I. Vinograd, S.-F. Wu, B. Vignolle, H. Mayaffre, S. Kr\''{a}mer, 
T. Kurosawa, N. Momono, M. Oda, J. Chang, C. Proust, M.-H. Julien, and D. LeBoeuf,
Phys. Rev. B {\bf 103}, 115133 (2021).

\bibitem{okamoto10}
 S. Okamoto, D. S\'{e}n\'{e}chal, M. Civelli, and A. -M. S. Tremblay,
 Phys. Rev. B {\bf 82}, 180511(R) (2010).

\bibitem{tallon01}
J. Tallon and J. Loram, Physica C \textbf{349} 53 (2001).

\bibitem{yang06}
K.-Y. Yang, T. M. Rice, and F.-C. Zhang,
Phys. Rev. B {\bf 73}, 174501 (2006).

\bibitem{kyung06}
B. Kyung, S. S. Kancharla, D. S\'{e}n\'{e}chal, A.-M. S. Tremblay,
M. Civelli, and G. Kotliar,
Phys. Rev. B {\bf 73}, 165114 (2006).

\bibitem{stanescu06}
T. D. Stanescu and G. Kotliar,
Phys. Rev. B {\bf 74}, 125110 (2006).

\bibitem{liebsch09}
A. Liebsch and N.-H. Tong,
Phys. Rev. B {\bf 80}, 165126 (2009).

\bibitem{sakai09}
S. Sakai, Y. Motome, and M. Imada,
Phys. Rev. Lett. {\bf 102}, 056404 (2009).

\bibitem{sakai15}
S. Sakai, M. Civelli, Y. Nomura, M. Imada,
Phys. Rev. B {\bf 92}, 180503(R) (2015).

\bibitem{wu18}
W. Wu, M. S. Scheurer, S. Chatterjee, S. Sachdev, A. Georges, and M. Ferrero,
Phys. Rev. X {\bf 8}, 021048 (2018).

\bibitem{norman98}
M. R. Norman, M. Randeria, H. Ding, and J. C. Campuzano,
Phys. Rev. B {\bf 57}, R11093 (1998).

\bibitem{norman07}
M. R. Norman, A. Kanigel, M. Randeria, U. Chatterjee, and J. C. Campuzano,
Phys. Rev. B {\bf 76}, 174501 (2007).

\bibitem{nagelthesis}
Peter Nagel, PhD thesis, Karlsruhe university (2001)

\bibitem{zimmermann03}
M. v. Zimmermann, J. R. Schneider, T. Frello, N. H. Andersen, J. Madsen, M. Kall, H. F. Poulsen, R. Liang, P. Dosanjh, and W. N. Hardy, Phys. Rev. B {\bf 68}, 104515 (2003)




\section*{Supplementary References}


\bibitem{luthi}
Lüthi, B. Physical Acoustics in the Solid State, Springer Series for Solid-State Sciences, Vol. 148 (Springer, Berlin, New York, 2005).


 \bibitem{nohara95}
Nohara, M.; Suzuki, T.; Maeno, Y.; Fujita, T.; Tanaka, I. and Kojima, H. Unconventional lattice stiffening in superconducting LSCO single crystals Phys. Rev. B \textbf{52} 570-580 (1995)

\bibitem{kraut93}
Kraut, O., Meingast, C., Brauchle, G., Claus, H., Erb, A., Müller-Vogt, G., and Wühl, H. Uniaxial pressure dependence of \Tc~of untwinned  YBa$_2$Cu$_3$O$_{\rm x}$ single crystals for x=6.5–7. Physica C: Superconductivity \textbf{205}, 139–146 (1993)

\bibitem{junod89}
A. Junod Physica C \textbf{162-164} 482 (1989)

\bibitem{wuhl91}
H. Wuhl Physica C \textbf{185-189} 482 (1991)

\bibitem{claus92}
H. Claus PhysicaC \textbf{198} 42 (1992)

\bibitem{loram01}
Loram, J.W., Luo, J., Cooper, J.R., Liang, W.Y., and Tallon, J.L.  Evidence on the pseudogap and condensate from the electronic specific heat. Journal of Physics and Chemistry of Solids \textbf{62} 59–64 (2001)

\bibitem{marcenat15}
Marcenat, C. et al. Calorimetric determination of the magnetic phase diagram of underdoped Ortho-II YBCO single crystals. Nat. Commun.  \textbf{6} 7927 (2015).


 \bibitem{lei93}
Lei, M.; Sarrao, J. L.; Visscher, W. M.; Bell, T. M.; Thompson, J. D.; Migliori, A.; Welp, U. W. and Veal, B. W. Elastic constants of a monocrystal of superconducting  YBa$_2$Cu$_3$O$_{7}$  Phys. Rev. B \textbf{47}  6154-6156 (1993)

 \bibitem{nagel00}
P. Nagel \emph{et al.}, Phys. Rev. Lett. \textbf{85} 2376 (2000)


\bibitem{nagelthesis}
Peter Nagel, PhD thesis, Kalrsruhe university (2001)

\bibitem{ando02}
Y. Ando \emph{et al.} Phys. Rev. Lett. \textbf{88} 137005 (2002)


\bibitem{meingast91}
C. Meingast Phys. Rev Lett. \textbf{67} 1634 (1991)


\bibitem{welp92}
U. Welp Phys. Rev. Lett. \textbf{69} 2130 (1992)


\bibitem{welp94}
U Welp Journal of Superconductivity \textbf{7} 159 (1994)

\bibitem{barber21}
MarK E. Barber \emph{et al.} preprint at arXiv:2101.02923 (2021)


%
%

\end{references}
\end{document}